\documentclass[12pt,preprint]{aastex}

\slugcomment{}

\shorttitle{Saving Phase Information}
\shortauthors{Schmitt et al.}

\begin{document}


\title{
Navy Prototype Optical Interferometer Imaging of Line Emission Regions of $\beta$ Lyrae
Using Differential Phase Referencing}


\author{H. R. Schmitt\altaffilmark{1,2,3}, T. A. Pauls\altaffilmark{1},
C. Tycner\altaffilmark{4}, J. T. Armstrong\altaffilmark{1},
R. T. Zavala\altaffilmark{5},
J. A. Benson\altaffilmark{5}, G. C. Gilbreath\altaffilmark{6},
R. B. Hindsley\altaffilmark{1}, D. J. Hutter\altaffilmark{5},
K. J. Johnston\altaffilmark{7}, A. M. Jorgensen\altaffilmark{8},
D. Mozurkewich\altaffilmark{9}}
\altaffiltext{1}{Naval Research Laboratory, Remote Sensing Division, Code 7215,
4555 Overlook Ave. SW, Washington, DC\,20375}
\altaffiltext{2}{Interferometrics, Inc, 13454 Sunrise Valley Drive, Suite 240, Herndon, VA\,20171}
\altaffiltext{3}{email:henrique.schmitt@nrl.navy.mil}
\altaffiltext{4}{Department of Physics, Central Michigan University, Mt. Pleasant, MI\,48859}
\altaffiltext{5}{U.S. Naval Observatory, Flagstaff Station, 10391 West Naval
Observatory Road, Flagstaff, AZ\,86001}
\altaffiltext{6}{Naval Research Laboratory, Free Space Photonics Communications Office,
Code 5505, 4555 Overlook Ave. SW, Washington, DC\,20375}
\altaffiltext{7}{US Naval Observatory, 3450 Massachusetts Avenue NW, Washington,
DC\,20392}
\altaffiltext{8}{New Mexico Institute of Mining and Technology, 801 Leroy Place,
Socorro, NM\,87801}
\altaffiltext{9}{Seabrook Engineering, 9310 Dubarry Road, Seabrook, MD\,20706}



\begin{abstract}
We present the results of an experiment to image the interacting
binary star $\beta$~Lyrae with data from the Navy Prototype Optical
Interferometer (NPOI), using a differential phase technique to correct
for the effects of the instrument and atmosphere on the interferometer
phases.  We take advantage of the fact that the visual primary of
$\beta$~Lyrae and the visibility calibrator we used are both nearly
unresolved and nearly centrally symmetric, and consequently have
interferometric phases near zero.  We used this property to detect and
correct for the effects of the instrument and atmosphere on the phases
of $\beta$~Lyrae and to obtain differential phases in the channel
containing the H$\alpha$ emission line.  Combining the
H$\alpha$-channel phases with information about the line strength, we
recovered complex visibilities and imaged the H$\alpha$ emission using
standard radio interferometry methods.  We
find that the results from our differential phase technique are
consistent with those obtained from a more-standard analysis using
squared visibilities ($V^2$'s).  Our images show the position of the
H$\alpha$ emitting regions relative to the continuum photocenter as a
function of orbital phase and indicate that the major axis of the
orbit is oriented along p.a.$=248.8\pm1.7^{\circ}$.  The orbit is
smaller than previously predicted, a discrepancy that can be
alleviated if we assume that the system is at a larger distance from
us, or that the contribution of the stellar continuum to the H$\alpha$
channel is larger than estimated.  Finally, we also detected a
differential phase signal in the channels containing He~I emission
lines at 587.6 and 706.5~nm, with orbital behavior different from that
of the H$\alpha$, indicating that it originates from a different part
of this interacting system.

\end{abstract}



\section{Introduction}

One of the major challenges in optical intererometry is recovering
images, since fringe visibility phases are badly corrupted by the
atmosphere on short timescales.  A few attempts have been made to
reconstruct images from the squared visibility ($V^2$), an unbiased
estimator of the visibility amplitude (e.g., \citealt{Baldwin96},
\citealt{Quirrenbach94}), but in most cases, images in optical interferometry
are produced from $V^2$ and closure phases (e.g., \citealt{Mon07}).  A
closure phase is the sum of visibility phases around a triangle of
interferometer array elements; the atmospheric turbulence effects
cancel in the sum, but only a fraction $1-2/N$ of the phase
information from an array of $N$ elements is recovered.  

The challenge in optical interferometric imaging is to recover as much
of the phase information as possible.  Here we describe the technique
we developed to obtain H$\alpha$ differential phases from NPOI
observations of the interacting binary star $\beta$ Lyrae. We were
able to recover complex visibilities of this source, and, for the
first time, obtain H$\alpha$ images using standard radio
interferometric reconstruction techniques.

Differential phases, also known as phase referencing (see
\citealt{Quirrenbach99}, \citealt{Mon03} for a discussion on the subject)
is an interesting and powerful technique. Here
one needs multiwavelength observations, obtained simultaneously, and a
priori knowledge about the structure of the source in a region of the
observed spectrum.  From that knowledge, one calculates the expected
phases in that spectral region, corrects for the atmospheric and
instrumental effects, and interpolates or extrapolates the corrections
to the wavelengths of interest.  One of the first attempts to use this
technique was presented by \cite{Vakili97,Vakili98}, who used
H$\alpha$ and He~I~667.8~nm GI2T observations of P~Cygni and
$\zeta$~Tauri to detect structures in the line-emitting regions of
these stars. However, they did not reconstruct images based on their
measurements; their conclusions were based only on the analysis of the
phases and visibilities of the stars. Another application of the
differential phase technique was presented by \cite{akeson00}, who
used the Keck Interferometer to detect variations of phase vs.\
wavelength in data from the binary star $\iota$~Peg.  An ideal class
of objects to which the differential phase technique can be applied is
Be stars, which are composed of a B star usually unresolved by
interferometric observations ($<1$~mas), surrounded by a disk of gas
with a diameter of a few mas, which is detected in H$\alpha$ line
emission (\citealt{Tycner05}).

The eclipsing binary star $\beta$ Lyrae (HD~174638, HR~7106, FK5~705) has
been extensively studied since it was discovered to be an eclipsing
system over 220 years ago
\citep{Goodricke1785}. Nevertheless, it remains one of the most baffling
binary systems known (see \citealt{Harmanec02} for a comprehensive
review). The current model consists of a $\sim$3 M$_{\odot}$ B6-8II star (star 2
in Harmanec's notation) which has filled its Roche lobe and is losing mass
at a rate of 2$\times10^{-5}$ M$_{\odot}$ year$^{-1}$ to a $\sim$13 M$_{\odot}$
early B star (star 1), which is completely embedded in and hidden by a thick
accretion disk. This accretion disk simulates a pseudophotosphere, which appears
to an observer to have a spectrum of an A5III star.
The orbit is very nearly circular, the orbital period is a
little over 12.9 days, and the period is increasing by 19 seconds per
year. The system lies at a distance of $270 \pm 39$~pc, as measured by
Hipparcos \citep{Perryman97}. The spectrum of $\beta$ Lyrae
contains at least six spectroscopic line components \citep{Harmanec02}. The three
major components are: (a) strong absorption lines from star 2 used to
derive the radial velocity curve having an amplitude of 370 km~s$^{-1}$,
(b) ``shell'' absorption components seen in the hydrogen, helium, and
many metallic lines, believed to arise from accretion disk, and (c) strong
lines of hydrogen and helium seen in emission which vary with orbital
phase and with time. One of the features that makes $\beta$ Lyrae unique
among contact binaries is the likelihood that the strong emission
lines come from a bipolar jet that has formed near the point where
material streaming from star 2 interacts with the accretion disk
around star 1 \citep{Harmanec96}.

This paper is organized in the following way. In Section 2 we present the
observations and reductions. In Section 3 we describe the corrections applied
to the data in order to obtain differential phases and describe the results
obtained for $\beta$ Lyrae. In Section~4 we describe the imaging of the H$\alpha$
emission of this source. In Section~5 we obtain information about the
orbit of this system from the displacement of the H$\alpha$ emission relative
to the continuum photocenter. In Section~6 we compare the results obtained from the
analysis of the H$\alpha$ results with those obtained using a traditional
V$^2$ analysis, and in Section~7 we give a summary of our results.

\section{Observations and Reductions}\label{sec:obs}

\subsection{Interferometry}

\begin{figure}
\plotone{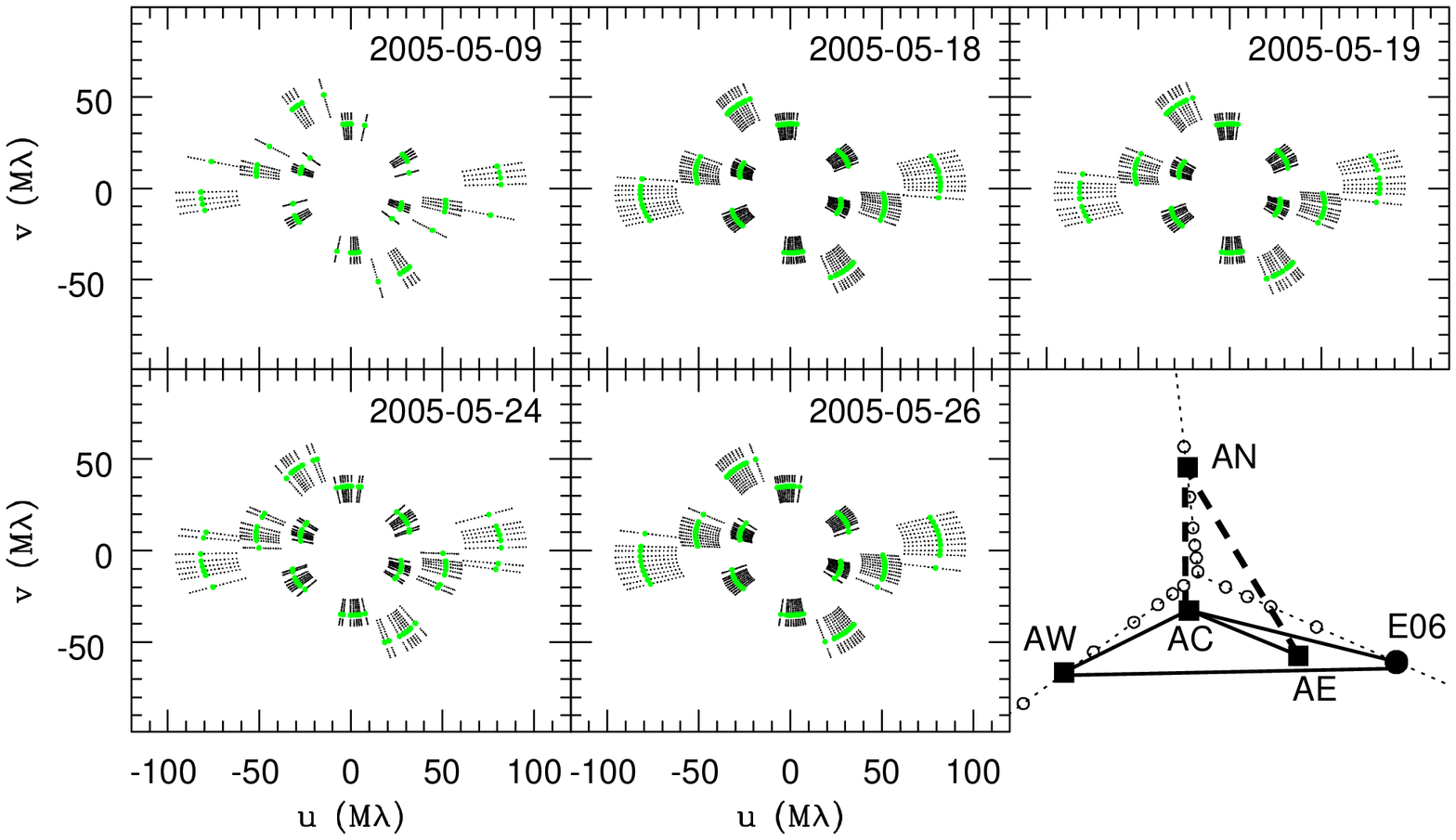}
\caption{The $u-v$ coverage of the different nights. The black points show the
continuum channels, while the green ones show H$\alpha$. The observing date is
presented in the top right corner of each panel. The bottom right corner of the
figure shows the array layout. The astrometric stations are shown as black squares
and E06 is shown as a black dot. We also show the baselines used all nights as
solid lines and the baselines that were used only on some nights, those involving
AN, as dashed lines. The maximum baseline lengths can be found in
Table~\ref{tbl-1}. \label{fig1}}
\end{figure}

Our observations were taken with the NPOI \citep{Armstrong98} on five different nights
between 2005 May 09 and 2005 May 26. The observations were made with two spectrographs,
simultaneously recording fringes in 16 spectral channels in the wavelength range
560--860~nm. Between one and three baselines were measured with each spectrograph.
The maximum baseline lengths ranged from
18.9 to 53.3~m. In Table~\ref{tbl-1} we present a log of the observations,
including the number of scans observed each night, the baselines used and the range
of hour angles covered by the observations. Figure~\ref{fig1} presents the
$(u,v)$-plane coverage achieved by the observations on each night. For reference,
this figure also shows the array layout with the stations and baselines used.
The observations of $\beta$ Lyrae
were interleaved with observations of a nearby calibrator star $\gamma$~Lyrae
(B9III, $V=3.24$~mag), which has a similar spectral type and magnitude to that of
$\beta$Lyrae. Each calibrator or source scan had a duration of 30~s.

Throughout this paper we will use orbital phases of $\beta$~Lyrae
based on the latest ephemeris of primary eclipses from \cite{Kreiner04}\footnote{
http://www.as.wsp.krakow.pl/ephem/LYR.HTM}.
We assumed a single orbital phase for
each night, corresponding to the midpoint of the observations (Table~\ref{tbl-1}).
This assumption will have no effect on the final results of this paper,
because our observations usually lasted for $\sim$2~hours, which corresponds
to $\sim$0.6\% of an orbital period. Comparing the values obtained from this
ephemeris with those from \cite{Harmanec93} we find that the two predictions
agree at the 2\% level, which is sufficient
for the purposes of the analysis presented here. As a final check,
we compared the predicted times of minima with visual measurements from
AAVSO\footnote{www.aavso.org},
spread over a period of 2 months around our observations. We used only validated
observations available at the AAVSO site and got an agreement better than 0.25
days between the observed and predicted minima, which corresponds to $\sim$2\%
of the orbital period.

The initial data reductions followed the incoherent and coherent integration
methods described in \cite{Peterson06a}. In the case of the incoherent
integration we start with data that were processed to produce squared visibilities
($V^2$) averaged into 1~s intervals. These data are then flagged to eliminate
points with fringe tracking and pointing problems \citep{Hummel98,Hummel03a}.
The flagged data set is then used to bias correct the $V^2$ data,
which are then calibrated.  The calibrated $V^2$ values are later used
to estimate the amplitudes of the complex visibilities.

For the coherent integration of complex visibilities, we use the
technique presented by \cite{Hummel03b} (see \citealt{Jorg07} for a
different algorithm).  We treat the visibilities in each 2~ms frame
(the instrumental integration time) as phasors.  To determine the
phase shift from one fame to the next, we compute the power spectrum
of the channeled visibility as a function of delay and measure the phase
difference between the peak of the fringe envelope and the nearest
fringe peak.  We rotate the phasors to correct for this residual phase
and average the complex visibilities in 200~ms subscans, which are used to
determine the visibility phases.
In Section 3, we describe other corrections applied to these data in
order to obtain H$\alpha$ differential phases for $\beta$~Lyrae.

\subsection{Spectroscopy}

\begin{figure}
\plotone{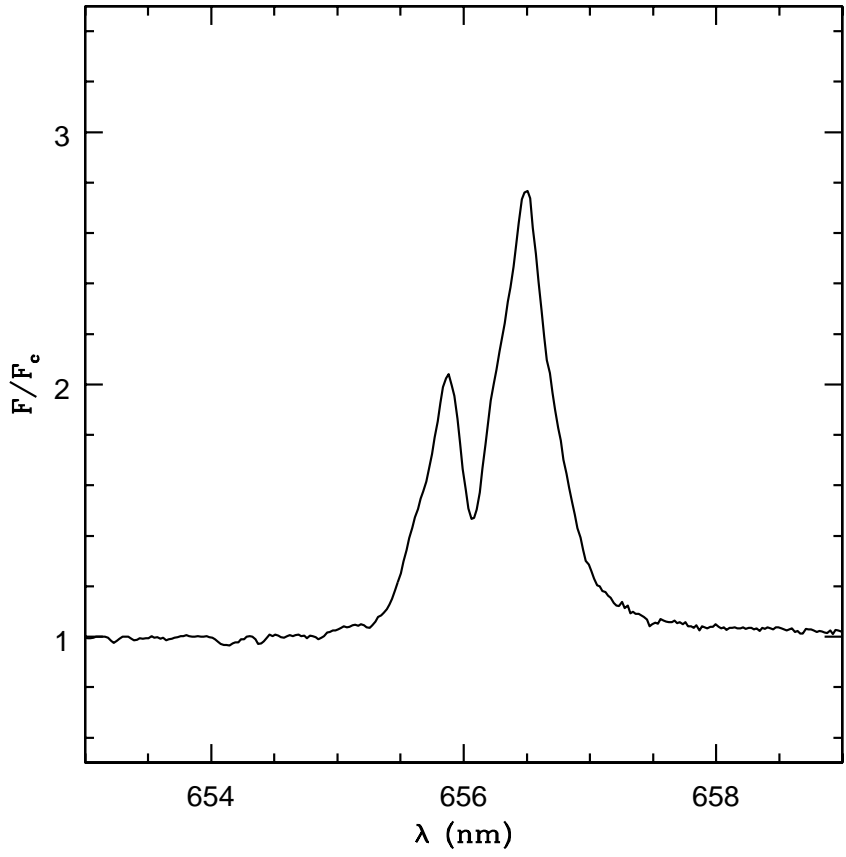}
\caption{Normalized high resolution H$\alpha$ profile of $\beta$ Lyrae. \label{fig2}}
\end{figure}

To estimate the strength of the H$\alpha$ emission in the $\beta$~Lyrae
system, we have obtained a high resolution spectrum 
using a fiber-fed Echelle spectrograph at the Lowell
Observatory's John S.\ Hall telescope.  The spectrum was acquired on
2005 May 19, which corresponds to the middle of our interferometric
run. The spectroscopic data were processed using standard routines
developed specifically for the instrument used in the
observations~\citep{Hall94}.  The final reduced spectrum in the
H$\alpha$ region reaches a resolving power of 10,000 and
signal-to-noise ratio of few hundred (see Fig.~\ref{fig2}).

We obtain an equivalent width (EW) of $-1.5 \pm 0.05$~nm for the H$\alpha$ profile
shown in Figure~\ref{fig2}.  Most of the uncertainty in the EW measurement is due
to the uncertainty associated with continuum normalization, which is
estimated to be of the order of 3\%.  To obtain the total emission in
the H$\alpha$ line we also need to correct for the underlying stellar
absorption line that has been filled in by the emission from the
circumstellar material.  We estimate the EW of the underlying
absorption line using the calibration developed by \cite{CW87}.  Using
a reddening corrected color $(B-V)_0=-0.05$~mag for $\beta$~Lyrae from
\cite{DP85}, we obtain an EW of 0.72~nm for the H$\alpha$ absorption
component, and therefore we estimate that the EW of the {\it total}
H$\alpha$ emission is $-$2.2~nm.  Considering that the NPOI
H$\alpha$ channel has a width of $15 \pm 1$~nm and the total emission line width
is $-2.2$~nm, we derive a fractional contribution from the
stellar photosphere~($c_{cnt}$) to the total flux in the channel of
$0.87 \pm 0.01$.

\section{Differential phases}

The challenge of recovering phase information from optical
interferometric observations results from atmospheric and instrumental
effects changing the phase calibration on a time scale of a few
milliseconds.  Fringe tracking can reduce these variations but not to
the level needed for good imaging.  Even with fringe tracking, the
solution starts with 1) short exposures that freeze the fringe motion,
2) software that measures and removes these variations and 3) a coherent
integration of hundreds of milliseconds (or more) needed to increase the
signal to noise to a level sufficient for the rest of the data
processing.  The result is measurements of the real and imaginary parts
of the visibility.  The uncorrected phase variations during the coherent
integration consist of a constant part and zero-mean fluctuations.  The
fluctuations reduce the measured visibility amplitude but have no effect
on the phase.  Thus these visibility data can be processed using closure
relationships or self-calibration.

However, with a small number of stations, there are significantly more
baselines than closure phases, so a significant fraction of phase
information is lost.  By contrast, the differential phase technique
described here provides good phase information for every baseline. The
idea is to recognize that the observed phase consists of three terms, viz.\
the desired source phase $\phi_0$, and offsets caused by the atmosphere
and the instrument:
\begin{equation}
\label{eqn:phi}
\phi_{obs} = \phi_0 + \phi_{atm} + \phi_{inst} ~.
\end{equation}

All terms have implicit wavelength dependencies.  The differential phase
technique consists of 1) measuring $\phi_{obs}$ at several wavelengths,
2) modeling the three right-hand-side terms in Equation~\ref{eqn:phi}
and 3) deriving parameters of the model by fitting it to the measured
$\phi_{obs}$.  For the NPOI, $\phi_{inst}$ is stable and can
be estimated by observing calibration stars; no parameters need to be
fit.  The atmospheric term, $\phi_{atm}$, is given by
\begin{equation}
\label{eqn:phi_atm}
\phi_{atm} = a(n-1)k + dk + \phi_c~,
\end{equation}
where $a$ is the differential air path between two beams, $n$ is the
refractive index of air, $k$ the wavenumber $(2\pi/\lambda)$, $d$ the
vacuum path difference and $\phi_c$ a wavelength independent phase
offset (Jorgensen et al., 2006).  Three quantities need to be fit:  $a$, $d$,
and $\phi_c$.  Thus with at least four spectral channels, some information
about the source can be obtained for {\it each} baseline.

\subsection{Instrumental and Atmospheric Corrections}

The H$\alpha$ emitting circumstellar disks of Be stars constitute one
of the best cases for which one can calculate instrumental and
atmospheric corrections to the visibility phases.  This correction is
based on observations of calibrator stars and the continuum channels
of the star of interest.  Since the photospheres of these stars usually do
not deviate significantly from point symmetry, we can model the
continuum as having zero intrinsic phase.  Even in the case of rapidly
rotating stars, in which the stellar image may deviate from point
symmetry, the phase should vary smoothly and slowly with wavelength 
\citep{Yoon07, Peterson06a, Peterson06b}.  The same principle applies
to calibration stars.  These properties allow one to use the
information from the continuum channels to calculate the appropriate
corrections to the phases and interpolate them over the channel
containing the H$\alpha$ line.

\begin{figure}
\plotone{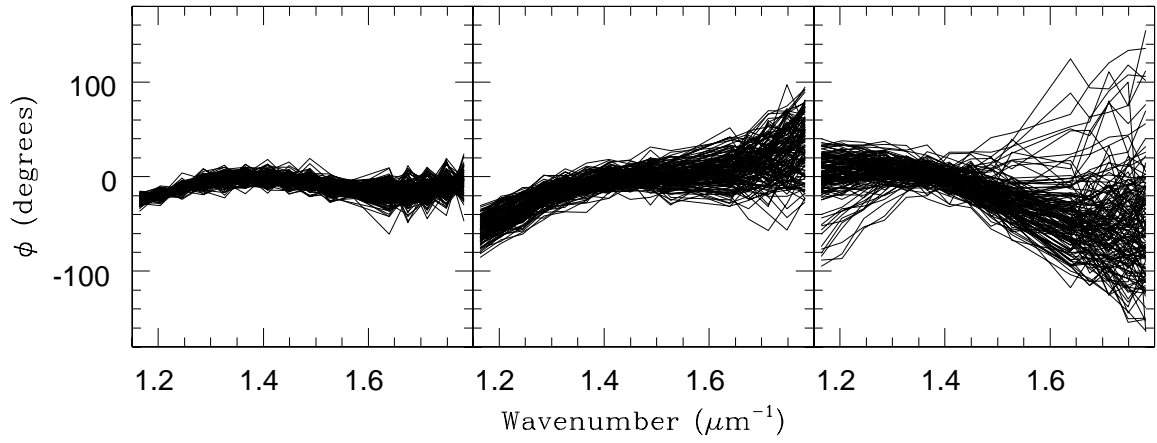}
\caption{Coherently integrated phases of 3 scans of $\gamma$~Lyrae, one per panel,
observed on 2005May09 with baseline AW0-E06.
Each line corresponds to a different 200~ms subscan.  \label{fig3}}
\end{figure}

A first example of the instrumental and atmospheric effects on the phases of a
partially resolved star is shown in Figure~\ref{fig3}. Each panel of this figure shows
a different scan of $\gamma$~Lyrae, while the different lines in each panel
represent individual 200~ms subscans. Notice that the lines of the different
subscans follow a 
similar cubic-like shape as a function of wavenumber, with some scatter. This cubic
curve can be explained as an instrumental phase, due to uncompensated amounts
of glass along the different beams. This curve does not change significantly from
scan to scan, or from night to night. A further inspection of this figure shows that,
besides this cubic component, one can also see a significant variation of
the phase among different subscans. These variations are due to the differential
air path between the two beams, and the shape of the curve is due to the fact
that $(n-1)$ has a significant dependence on $k^2$ \citep{owens67}. One can also
see in Figure~\ref{fig3} that the scatter is much smaller in the left panel, during
the first scan of the night, increasing significantly toward later times (right),
consistent with the worsening of the seeing conditions through the night.

In order to obtain differential phases of Be stars, one needs to correct the data for
the two components described above. For the instrumental phases we use the values
determined by averaging all the 200~ms subscans of calibrator stars observed throughout
the night. Since the effects of the differential air path and vacuum delay are
additive, one can reasonably assume that they should average down to zero
over the course of a night. Also, since calibration stars are expected to have
zero phase in the spectral range covered by our observations, by averaging the
complex visibilities of the calibrators one is left only with instrumental
phases, which can then be subtracted from the program star visibilities. 

\begin{figure}
\plotone{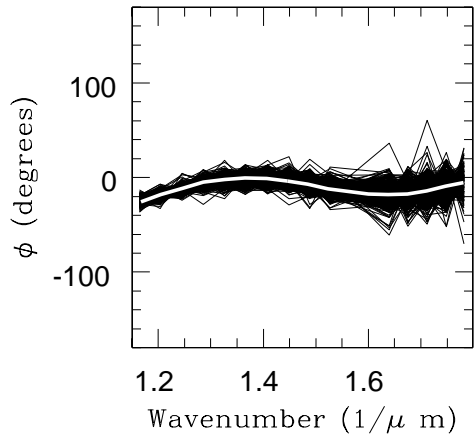}
\caption{Comparison between the phases of individual subscans from Figure~3, corrected
for atmospheric and vacuum delay effects, and the instrumental phase (white line),
obtained by averaging all subscans of the calibrator star observed in this particular
night. The uncertainty in the instrumental phase is 0.15$^{\circ}$ in the red
part of the spectrum, growing to $\sim0.5^{\circ}$ in the blue.\label{fig4}}
\end{figure}

We demonstrate the procedure used to obtain instrumental phases in Figure~\ref{fig4}.
Here we show the phases obtained by averaging the visibilities of all subscans
of $\gamma$ Lyrae observed on a night, 720 in total. We can also show that the
other component affecting the phases of the subscans is due to differential
air path between the two beams. Following the derivation presented by
\cite{Armstrong98}, we expand Eq.~\ref{eqn:phi_atm} into a power series around $k_0$ and
rewrite Eq.~\ref{eqn:phi} in the following form:
\begin{equation}
\label{eqn:phi_expanded}
\phi_{obs} = \phi_0 + \phi_{inst} + c_0 + c_1 k + c_2 k^2 .
\end{equation}
Starting from this equation and assuming that $\phi_0=0$, one can then solve for
the coefficients $c_0$, $c_1$ and $c_2$ for each subscan, such that the residuals
produce the best fit to the atmosphere phases ($\phi_{obs} - \phi_{inst}$).
We show in Figure~\ref{fig4}, as black lines, all subscans after subtracting only the
atmosphere effects from the observed phases.
Notice that these lines follow the instrumental one, with
an r.m.s. of 0.15$^{\circ}$ in the red and $\sim$0.5$^{\circ}$ in the blue.

\subsection{Differential Phases of $\beta$ Lyrae}

\begin{figure}
\plotone{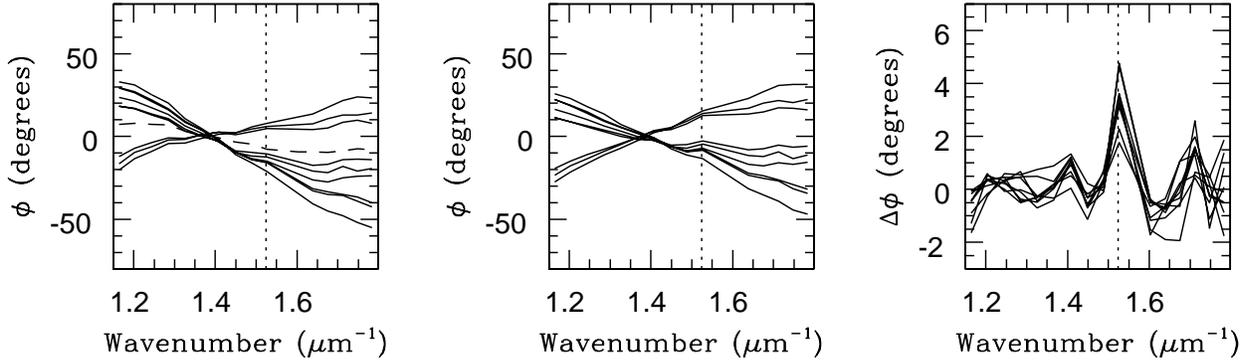}
\caption{The different correction steps applied to the data of $\beta$ Lyrae.
The first panel (left) shows the average phases of each
scan (solid lines) as for baseline E06-AC0, obtained on 2005May19. The dashed
line shows the instrumental phases and the dotted vertical line shows the
location of the H$\alpha$ channel. The middle panel shows the phases
obtained after subtracting the instrumental contribution. The right panel
shows the differential phases obtained after subtracting a quadratic function,
which was fitted to the continuum points only. The uncertainty in the phases
is of the order of 0.3$^{\circ}$ in the red, 0.5$^{\circ}$ at H$\alpha$,
growing to 1$^{\circ}$ in the blue. \label{fig5}}
\end{figure}

The correction steps applied to the data of the program star ($\beta$ Lyrae)
to obtain the phase in the H$\alpha$ channel are presented in Figure~\ref{fig5}.
In order to simplify the procedures and reduce the number of calculations,
we take advantage of the fact that the atmospheric effects constitute an
additive phase term, so instead of dealing with individual 200~ms
subscans we can work with the average of all visibilities in a scan. The left
panel of Figure~\ref{fig5} shows the average phases of nine scans of $\beta$~Lyrae.
In the central panel we show the phases of the individual scans after the
subtraction of the instrumental phase, shown as a dashed line in the left panel.
The last step of the process is the correction of the atmospheric effects,
by fitting a quadratic function of the form of
Eq.~\ref{eqn:phi_expanded} to the continuum channels for each average scan
and subtracting the resulting function from all the channels.

In the procedures described above, the continuum phases of
$\beta$~Lyrae were assumed to be zero.  Nevertheless, one should
expect to see phase variations across the spectrum.  $\beta$~Lyrae is
a binary system with a maximum separation of $\sim 1$~mas and $\Delta
m \sim 0.9$~mag, with almost no color difference between the two
components \citep{Wilson74, Harmanec02}.  Using Eq.~8 of
\cite{Mourard92}, we calculate that this system, observed at maximum
separation with a 50~m baseline aligned with the separation vector,
should show a continuum phase $\phi_{cnt}$ at the H$\alpha$ channel of
$\sim 14^{\circ}$, with a slope of $1^{\circ}$ per channel.  This case
corresponds to the longest baseline in our data, but most of our
baselines were 34~m long or less.  For a 30~m baseline, $\phi_{cnt}$
at H$\alpha$ is $1.6^{\circ}$ with a slope of $0.2^{\circ}$ per
channel.

Although the continuum phase component varies smoothly with
wavenumber and can be modeled by our atmospheric phase subtraction method,
arbitrarily setting this component to zero can in principle influence our results.
The complex visibility of the H$\alpha$ channel results from the sum of the
complex visibility of the continuum and H$\alpha$ emission line. Setting the
continuum phase to zero will change the phase of the emission line visibility
by $\phi_{cnt}$. For baselines of 30~m or smaller this is not an important
effect, given the small continuum phases. However, as shown above, in the case of
our longest baseline this can change the H$\alpha$ emission phase by as much as
$\sim14^{\circ}$. We discuss in Section~4 how we tested the effect of this
assumption on the imaging of this source.

\begin{figure}
\plotone{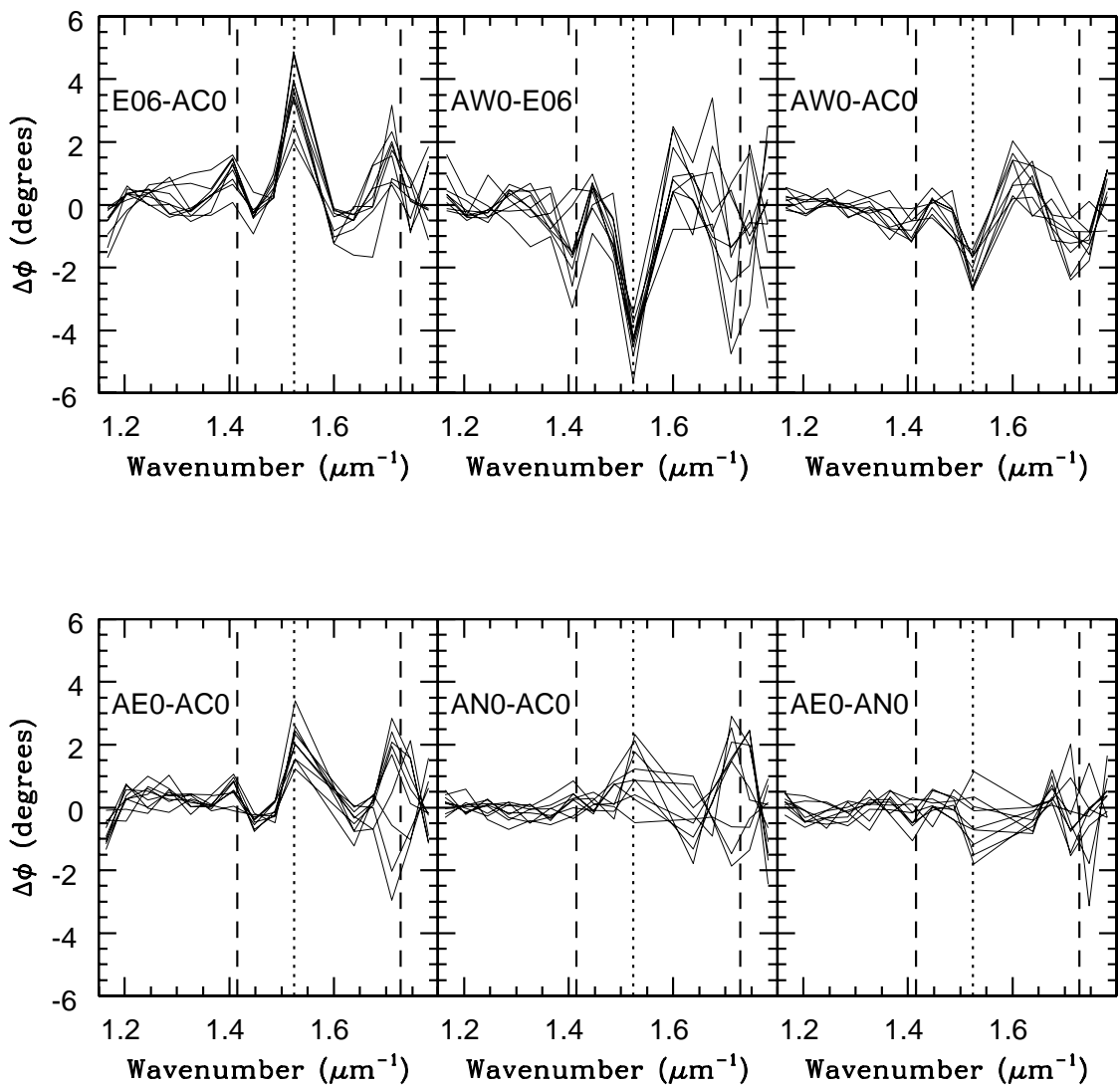}
\caption{Residual phases of $\beta$ Lyrae after correcting for the instrumental,
atmospheric and vacuum differential delay effects. Each panel shows a different
baseline, observed on 2005May19. The individual lines represent individual
scan averages. The vertical dotted line indicates the wavenumber of
H$\alpha$, while the dashed ones show the wavenumbers of HeI. \label{fig6}}
\end{figure}

Figure~\ref{fig6} shows the differential phases of $\beta$~Lyrae on all baselines
observed on 2005 May 19. Here we see results similar to those from Figure~\ref{fig5},
with the continuum phases around zero, and the H$\alpha$ and He~I channels showing
a differential phase. The strongest H$\alpha$ phase is found on
baselines with the longest extent along the E-W direction, with very little
differential phase signals on baselines AN0-AC0 and AE0-AN0. Comparing
these results with the ones from
\cite{Mourard92} and \cite{Harmanec96} we get a consistent picture for
this system. Based on GI2T observations, \cite{Harmanec96} 
concluded that the orbit of the binary component should be roughly oriented
along the E-W direction, since their observations used a N-S baseline and did
not resolve the two stars. On the other hand, their line observations detected
smaller $V^2$'s for H$\alpha$ and He~I indicating that the line emitting region is
more extended along the N-S direction. This result is consistent with the
spectropolarimetric observations from \citep{Hoffman98}, who found that
H$\alpha$, He~I 587.6~nm and ultraviolet ($\lambda<360$nm) emission are
polarized along p.a. $\sim74^{\circ}$. This result indicates that this emission is
polarized by a structure perpendicular to this direction.
Another result that confirms this geometry is a radio nebula
along p.a. $156 \pm 4^{\circ}$ detected by \cite{Umana00}.

Comparing our results to those from \cite{Harmanec96} suggests a possible
contradiction, since we do not see a strong differential phase on the
baselines oriented closer to the N-S direction, at least when comparing
their signal to that measured on baselines oriented closer to the E-W direction.
However, our observations can be explained in a way that agrees with the previous
results. First, the weaker phase signal along the N-S direction can be explained
in part as an effect of the baseline lengths of the two observations.
\cite{Harmanec96} used a baseline of 51~m, while for our AE0-AN0 baseline
we have a longest projected length of only $\sim 25$~m along the N-S direction.
Our maximum baseline length along the E-W direction is roughly twice that, which
can explain the stronger differential phase signal along this direction.
Second, we interpret the differential phases as a displacement between the 
H$\alpha$ line emitting
region and the continuum photocenters. Since the binary system is expected
to be oriented along the E-W direction, this would explain the stronger H$\alpha$
differential phases on baselines aligned closer to this direction.

\begin{figure}
\plotone{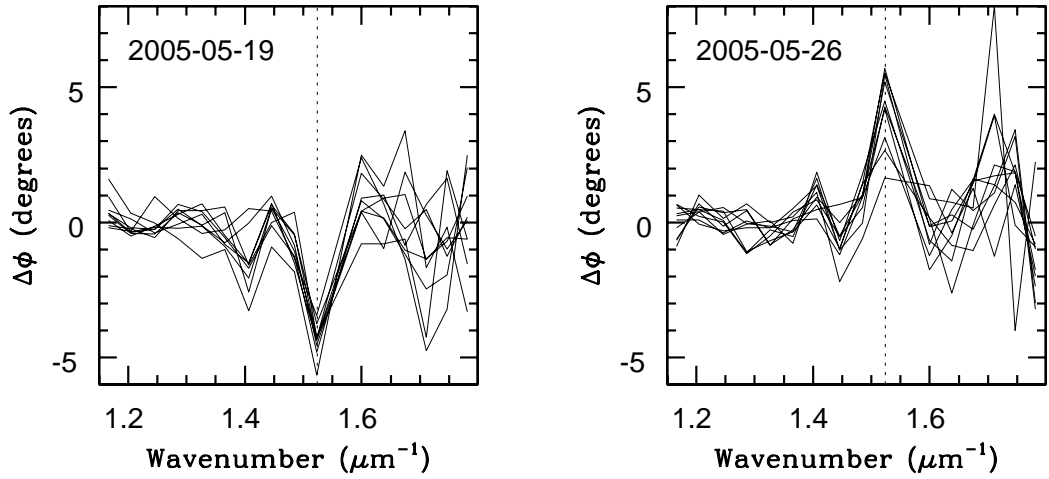}
\caption{Comparison of the differential phases of the longest baseline AW0-E06
(53~m) on the nights of 2005May19 (left) and 2005May26 (right). The two
observing nights are separated by approximately half an orbital period.
\label{fig7}}
\end{figure}

Further support for this interpretation is presented in Figure~\ref{fig7}, where we 
show the differential phases of baseline AW0-E06 on 2005 May 19 and
2005 May 26. These observations correspond to orbital phases of 0.24 and 0.78,
respectively. Here we can see that the H$\alpha$ phase signal changes from
negative to positive between the two dates, which can be understood as
the center of line emission and continuum photocenter changing orientation
relative to each other.

We would also like to point out another interesting result from Figure~\ref{fig6},
the variation of the He~I differential phases near $k \sim 1.4$ and $1.7~\mu$m$^{-1}$,
in particular on baselines AE0-AC0 and AN0-AC0. One can see in this figure
that the He~I channel phases vary significantly from scan to scan, with the sign
of the phase changing from positive to negative in a few scans. When we
compare this behavior to that of the H$\alpha$ phases we see that H$\alpha$
does not change sign on the same night. This result indicates that the He~I
emitting region has a structure different from that of the H$\alpha$ region.
This conclusion is supported by the spectropolarimetric results from \cite{Hoffman98},
who found that He~I 587.6~nm and 706.5~nm have a different polarization properties
from those of H$\alpha$ and He~I~667.8~nm. This difference is due to the way these
lines are excited, with the 587.6 ($k = 1.70~\mu$m$^{-1}$) and 706.5~nm
($k = 1.42~\mu$m$^{-1}$)
being produced in higher density regions, possibly in the inner accretion disk.

\section{Imaging}

The ultimate goal of our differential phase experiment is the recovery of
complex visibilities of this star in order to image the H$\alpha$ and
continuum emission. In order to do this we start by combining the final phase
information described above with the calibrated $V^2$'s from the incoherent
integration method to calculate the corresponding complex visibilities
for both continuum and line channels. 
Since these visibilities have been corrected for instrumental and atmospheric
effects, the continuum channels can be exported into a FITS file and imaged
using the usual radio interferometry techniques. However, in order
to image the H$\alpha$ emission one needs to take into account the fact that
most of the light in this channel originates from the continuum.

The correction for the continuum contribution to the H$\alpha$ channel is
calculated using the $c_{cnt}$ value obtained from the spectroscopic measurements.
Rewriting Eq.~1 of \cite{Tycner06} in a more general form, we have the following
relation for the continuum subtracted H$\alpha$ complex visibility:
\begin{equation}
\label{eqn:V_Halpha}
V_{{\rm H}\alpha} = (V_{obs} - c_{cnt} V_{cnt}) / (1 - c_{cnt}) .
\end{equation}
$V_{obs}$ is the observed complex visibility in the H$\alpha$ channel, including
the line and continuum components, $V_{cnt}$ and $V_{{\rm H}\alpha}$ are the complex
visibilities of the continuum and H$\alpha$ components, and $c_{cnt}$ is the
fractional contribution from the stellar continuum to the total flux in
the H$\alpha$ channel. We use the average of the two channels adjacent to H$\alpha$
to estimate $V_{cnt}$ at the H$\alpha$ channel.

\begin{figure}
\epsscale{1.0}
\plotone{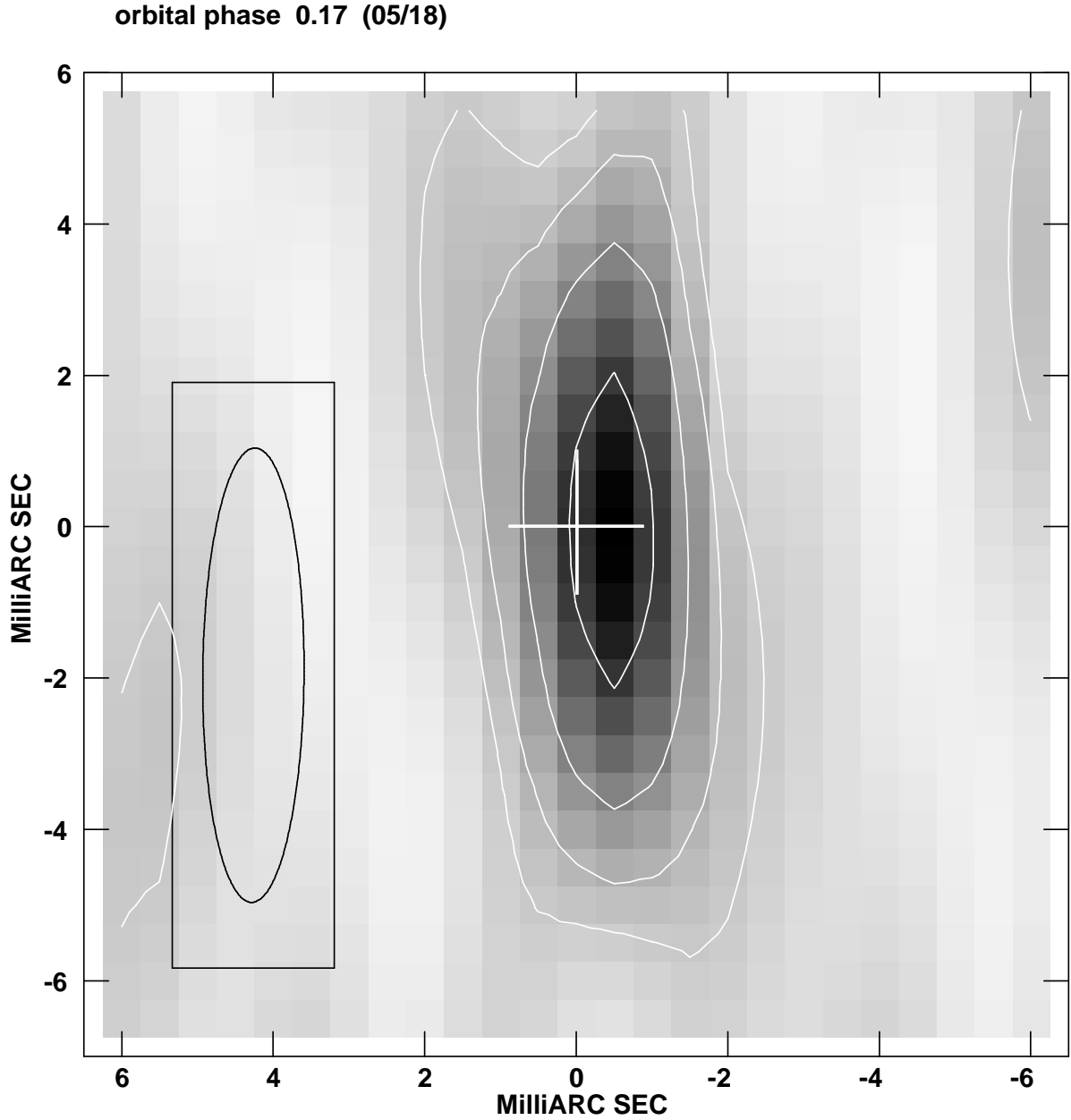}
\caption[]{The following 5 figures show the H$\alpha$ images of $\beta$ Lyrae.
Each figure corresponds to a different night and are organized in order of increasing
orbital phase, indicated at the top of the panel. The cross shows the photocenter
of the system, measured on the continuum images. The lowest contour corresponds
to the 3$\sigma$ level. The reconstructing beam size and orientation is shown in
the bottom left corner of each panel.
\label{fig8}}
\end{figure}

The resulting visibilities are converted into FITS files (one per night)
and imaged using AIPS\footnote{http://www.aoc.nrao.edu/aips/}
\citep{vanmoorsel96}.
The resulting H$\alpha$ images are presented in Figure~\ref{fig8}.
These images were created using pixel sizes of 0.5~mas and a hybrid weighting
scheme between natural and uniform weights, with robustness index 0.
Just for reference we show in each panel, as a white cross, the position of
the continuum photocenter, measured on continuum images restored from data
observed on the same night. The position of the continuum was always in the
center of the image, consistent with the fact that the continuum phases were
set to $\sim 0^{\circ}$ in our procedures. Notice that in most cases the H$\alpha$
emission has a shape similar to that of the restoring beam, which indicates
that  the emission is unresolved, or just partially resolved, with a size comparable
to that of the beam.

The most important result to be taken from Figure~\ref{fig8} comes from the comparison
between the peaks of the H$\alpha$ and continuum emission. We can see in the different
panels that these two points do not coincide, and that the displacement
of the H$\alpha$ photocenter changes relative to that of the continuum as
a function of orbital phase. Since the H$\alpha$ emission originates in the
disk around the more massive star, while the continuum photocenter is located
closer to the less massive one \citep{Harmanec96}, we are effectively imaging
the orbit of this system.

We also tested the effect introduced in the H$\alpha$ images of assuming that the
continuum phase on the longest baseline is zero. To solve for this effect,  we added 
14$^{\circ}$ to the phases of both H$\alpha$ and adjacent continuum channels of the
longest baseline before subtracting the continuum contribution to the H$\alpha$ 
channel. We did not try to apply this correction to other baselines because of the
small magnitude of their continuum phases. We followed the procedure described
above and obtained new images using the uncorrected short baseline H$\alpha$
visibilities and the corrected long baseline H$\alpha$ visibilities. We do not
find a significant change relative to the images obtained setting the continuum
phase to zero. The position of the peaks of emission changed less than
$0.4 \sigma$, indicating that setting the continuum phases to zero was
a reasonable approximation in this case.

\section{The Orbit of $\beta$ Lyrae}

Although we do not resolve the stars in this binary, our results can
be used to derive some properties of the $\beta$~Lyrae system. First, since
this is an eclipsing binary, it is straightforward to determine the
orientation of the system on the plane of the sky. The inclination of the
orbit is $86^{\circ}$ \citep{Linnell98, Linnell00} and its projection
on the sky can be
reasonably well approximated by a straight line. In Figure~\ref{fig9} we show the
position of the H$\alpha$ emission relative to the continuum
photocenter for the five  nights
of data used in this paper. These positions were obtained by fitting gaussians
to the images, and the values are given in Table~\ref{tbl-2}. 

Figure~\ref{fig9} also shows the orbit of the system, obtained by fitting a line
with instrumental weighting ($1/\sigma^2$) to the data points. We find that
the orbit is oriented along p.a.$= 248.8\pm1.7^{\circ}$. Notice that the point
from 2005 May 18 has a much higher uncertainty in the N-S direction and
deviates a little from the line. The uncertainty is caused by the fact that station AN0
was not used on this night, resulting in poorer resolution along this
direction. We also point out that the fitted line does not cross the origin,
which can be attribute to the fact that our observations cover only one half
of the orbit, from orbital phases 0.17 to 0.78. 

The reliability of the orbital orientation we obtain for the $\beta$~Lyrae
system can be checked through the comparison with other measurements available
in the literature. The radio observations presented by \cite{Umana00} detected
a jet-like nebular structure oriented along p.a.$= 156.5 \pm 4^{\circ}$. This nebula
is perpendicular, within uncertainties, to the orbit p.a. derived here, agreeing
with the expectation that jets are launched perpendicular to their
disks. Further evidence confirming the orbital orientation of this system
comes from the comparison of our results with the spectropolarimetric results
of \cite{Hoffman98}. They predict that the jets should be oriented along
p.a.$\sim 163.5^{\circ}$, similar to that found by \cite{Umana00} and
roughly perpendicular to the orbital orientation we found.

Finally, we compare our measurements of the displacement between the H$\alpha$ 
and continuum photocenters to predictions based on parameters of the
system derived by others (e.g.,
\citealt{Harmanec96, Harmanec02, Linnell00}). We use the same reference
system as that of \cite{Harmanec96}, where the orbit is in the $X$-$Y$ plane,
with the origin at the most massive star (the one which is
currently gaining mass) and the $X$-axis is defined by the line joining the
two stars. The analysis of the radial velocities of this system shows that
the eccentricity of the orbit is 0 \citep{Harmanec93}. According to \cite{Harmanec02},
the separation between the centers of the two stars is $58.52 R_{\odot}$.
For $\Delta m = 0.9$~mag the position of the photocenter is at
$X = 40.64 R_{\odot}$ and $Y = 0$. Based on the analysis of the velocity curve
of the H$\alpha$ emission \cite{Harmanec96} found that this emission line originates
in the disk ($X = 4.12 R_{\odot}$ and $Y = 4.97 R_{\odot}$), in a region offset
from the center of the system, possibly related to the point of impact of the gas
in the disk.

We used these parameters to calculate the orbits of the H$\alpha$ and continuum
photocenters, and converted them to positions in the sky using the inclination
derived by \cite{Linnell98} ($i = 86^{\circ}$), the orientation derived by us
(p.a.$= 248.8^{\circ}$) and a distance of $270 \pm 39$~pc. The resulting H$\alpha$
orbit relative to the continuum photocenter is shown in Figure~\ref{fig10}. Here we can see
that the orbit has the correct orientation, but has slightly larger semimajor
axis than the one measured by us. The maximum calculated separation between
the H$\alpha$ and continuum is 0.623~mas, which is $\sim 34$\% larger than the
value we measured (0.466~mas at orbital phase 0.24).

The difference between the observed and predicted  orbital semimajor axis
can be due to several effects. First, the uncertainty in the distance of this
system is fairly large (39~pc). Increasing the distance of this source to 309~pc
(corresponding to one standard deviation in the distance), the
apparent size of the
orbit is reduced by 14\%. To illustrate this effect we show in Figure~\ref{fig10}
(as a dotted line) the  orbit obtained using a distance of 309~pc. The uncertainty
in the $\Delta m$ of this system can also introduce errors in the size
of the orbit. If we assume $\Delta m = 0.8$mag instead of
0.9~mag, we get that the photocenter is at $X = 39.48 R_{\odot}$ and $Y = 0$,
which reduces the orbit size by $\sim 3$\%.

Another important source
of error in our measurements is the correction of the continuum contribution
to the H$\alpha$ channel. Factors that contribute to errors in this
correction are the uncertainty in the width of the channel and the EW of
the H$\alpha$ absorption component. In this paper we used the \cite{CW87}
calibration to correct for the underlying H$\alpha$ absorption. However,
their calibration is based on main sequence stars and may be
overestimating the correction for $\beta$~Lyrae, where the brightest
component is a B6-8II star.
We tested this hypothesis by using the extreme assumption that the
EW correction for the underlying absorption to H$\alpha$ is 0, which gives
$c_{cnt}=0.91$. Using this value, we obtained new images and measured
the position of H$\alpha$ emission relative to the continuum photocenter.
The new measurements are shown in Figure~\ref{fig11}, where we can see a result
opposite to the one in Figure~\ref{fig10}. In this case the observed semi major
axis is 0.8~mas at orbital phase 0.24, $\sim 28$\%, or larger than the predicted value.


\section{Comparison Between Differential Phase and $V^2$ Results} 

As a way to check the results obtained with the differential phase technique,
we use the calibrated continuum $V^2$ and closure phases to obtain
information about the binary system. These measurements are used to compute
the separation $\rho$ and position angle $\theta$ of the binary for each
night. The best fit values, which were obtained for $\Delta m(670~nm)=0.9$ mag
and assumed to be the same at other wavelengths, are given in
Table~\ref{tbl-3} and plotted in Figure~\ref{fig12}. This figure shows that
the maximum separation between the two stars is of the order of $\sim 0.9$~mas,
although with large uncertainties. Part of these uncertainties are related to
the fact that our observations do not fully resolve this system, as can be
seen in Figure~\ref{fig13}. Here we present, for each night, the $V^2$'s
of one scan observed with the longest baseline (AW0-E06, 53~m), and the 
corresponding best fitting models. We selected the
scans within 30 minutes of meridian crossing, which is close to
the highest resolution achieved with this baseline. This figure shows some of
the limitations of our measurements. We observe higher $V^2$'s on the night of
May 9, when the system is at orbital phase 0.47, and lower $V^2$'s on the other
nights, when the stars are closer to their maximum elongation. However, we do not
see the characteristic cosine wave signature of a binary, which makes the
determination of the separation of the two stars uncertain.

When comparing the results obtained from the V$^2$ analysis with the H$\alpha$
ones, we should keep in mind that the former correspond to the separation
between the photocenters of the two stars, while the latter correspond
to the separation between the H$\alpha$ and continuum photocenters. Consequently,
we need to convert the H$\alpha$ measurements to the same reference frame as
the continuum measurements. Following the
discussion presented in Section~5 we have that the two stars are separated
by 58.52~R$_{\odot}$, the continuum photocenter is located at 40.64~R$_{\odot}$
from the most massive star, calculated assuming $\Delta$m=0.9~mag, and for
simplicity we can say that the H$\alpha$ photocenter is located at
4.12~R$_{\odot}$ from this star. These numbers indicate that the measured
H$\alpha$ separation relative to the continuum photocenter at orbital
phase ~0.25 corresponds to only 62\% of the real separation between the
photocenters of the two stars.  Converting the measurements presented in
Figures~\ref{fig10} and \ref{fig11}, we have that the observed H$\alpha$
orbits correspond to separations in the range 0.75 to 1.28~mas between the
two stars (the range of values corresponds to $c_{cnt}=$0.87 and 0.91,
respectively). Considering the uncertainties, these values are in very
good agreement with those from the V$^2$ analysis presented in Figure~\ref{fig12}.

Another way to confirm the H$\alpha$ results is by comparing
the orientation of the orbits derived from the two
methods. A least squares fit to the data presented in Figure~\ref{fig12},
using instrumental weighting, gives a major axis orientation along
p.a.$=256.1^{\circ}\pm17.7^{\circ}$. This value is slightly different from
the one obtained from H$\alpha$ measurements, p.a. $=248.8^{\circ}\pm1.7^{\circ}$,
but they are still in agreement, given the uncertainties.

As a last check we used the method described by \cite{Tycner05, Tycner06}
to determine the size of the H$\alpha$ disk. We used the calibrated H$\alpha$
V$^2$'s, corrected to eliminated the continuum contribution from the binary star
to this channel. Following the results from \cite{Tycner06} we decided to fit
the H$\alpha$ V$^2$'s with a gaussian model, which results in a disk with
HWHM=0.60$\pm$0.10~mas (Figure~\ref{fig14}). The gaussian HWHM is comparable
to the Roche lobe radius of $\sim$0.52~mas (30.3~R$_{\odot}$ \citealt{Harmanec02}),
and consistent with the orbit obtained using the two methods described above.

\section{Summary}

In this paper we presented the development of a differential phase technique
using NPOI data. We described the methods used to correct the instrumental and
atmospheric effects on the phases of $\beta$ Lyrae, which allowed us to recover
the complex visibility of the H$\alpha$ channel. We also described the procedure
used to correct the continuum contribution to the H$\alpha$ channel and successfully
imaged the line emission of this system using standard interferometric techniques.
We found that the major axis is oriented along p.a. 248.8$^{\circ}$, consistent
with previous spectropolarimetric measurements, radio and optical interferometry
results. From the comparison between the observed position of H$\alpha$ relative
to the continuum photocenter and values obtained from models, we find that our
measurements indicate a semi major axis smaller than that predicted by the models.
We suggest that the most likely
causes for this discrepancy are a wrong distance, brightness ratio between the
2 stars and the uncertainty in the correction of the continuum contribution
to the H$\alpha$ channel. We also found that the results obtained using the
technique developed here are consistent with those obtained from a more
traditional analysis that uses only V$^2$ measurements.

The technique presented here represents a major step in the process of obtaining
images from optical interferometric observations. Due to several limitation, the most
important being the rapid time scales in which the atmosphere changes, the phases
of a source wrap around too fast. Consequently, this information is lost and 
one cannot apply the usual image reconstruction techniques employed in
radio interferometry. By making some assumptions about the structure of the source,
our technique allows us to recover this information and
obtain images for the emission line channels. In the case of $\beta$ Lyrae this
allowed us to measure the orientation and size of the orbit using the relative
location of the H$\alpha$ and continuum photocenters. We also found that the 
HeI emission has a different structure from that of H$\alpha$.

Besides $\beta$ Lyrae we are also applying this technique to other Be systems.
This will allow us to image their circumstellar disk, map structures and study
in better detail the accretion flow and effects such as the ionization of the
disk by a binary companion.

\acknowledgments

The work done with the NPOI was performed through a collaboration between
the Naval Research Lab and the US Naval Observatory, in association with
Lowell Observatory, and was funded by the Office of Naval Research and the
Oceanographer of the Navy. We thank the NPOI staff for the careful
observations that contributed to this work. C. T. thanks Lowell Observatory
for the generous time allocation on the John S. Hall 1.1 m telescope and
thanks Wes Lockwood and Jeffrey Hall for supporting the Be star project
on the Solar-Stellar Spectrograph. This research has made use of the
SIMBAD literature database, operated at CDS, Strasbourg, France. We
acknowledge with thanks the variable star observations from the AAVSO
International Database contributed by observers worldwide and used in this research.
We would like to thank Christian Hummel for the availability of the 
data reduction package OYSTER.


{\it Facilities:} \facility{NPOI}, \facility{USNO}, \facility{Lowell}.

\clearpage

\begin{figure}
\plotone{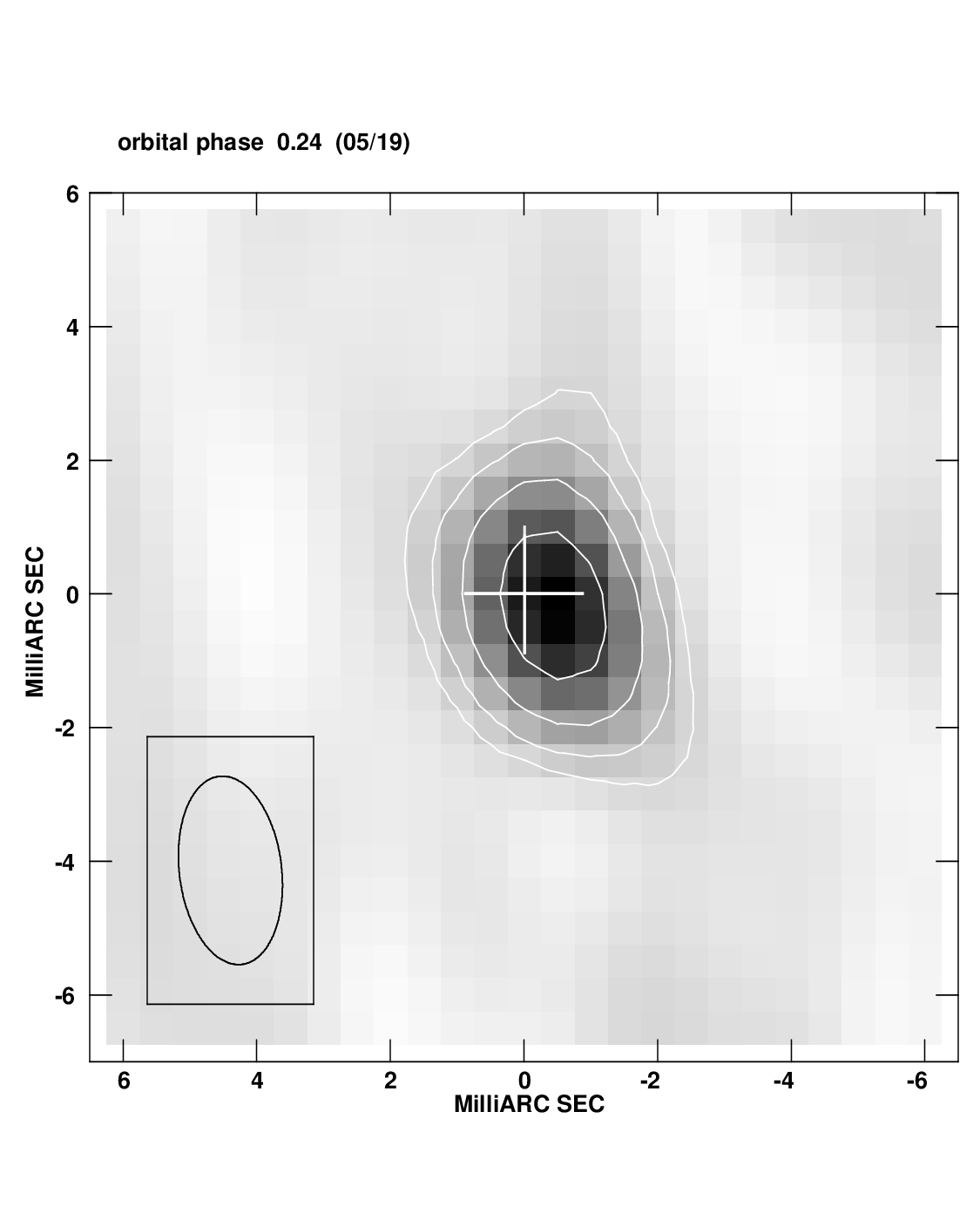}
\end{figure}

\begin{figure}
\plotone{f8c.eps}
\end{figure}

\begin{figure}
\plotone{f8d.eps}
\end{figure}

\begin{figure}
\plotone{f8e.eps}
\end{figure}

\begin{figure}
\plotone{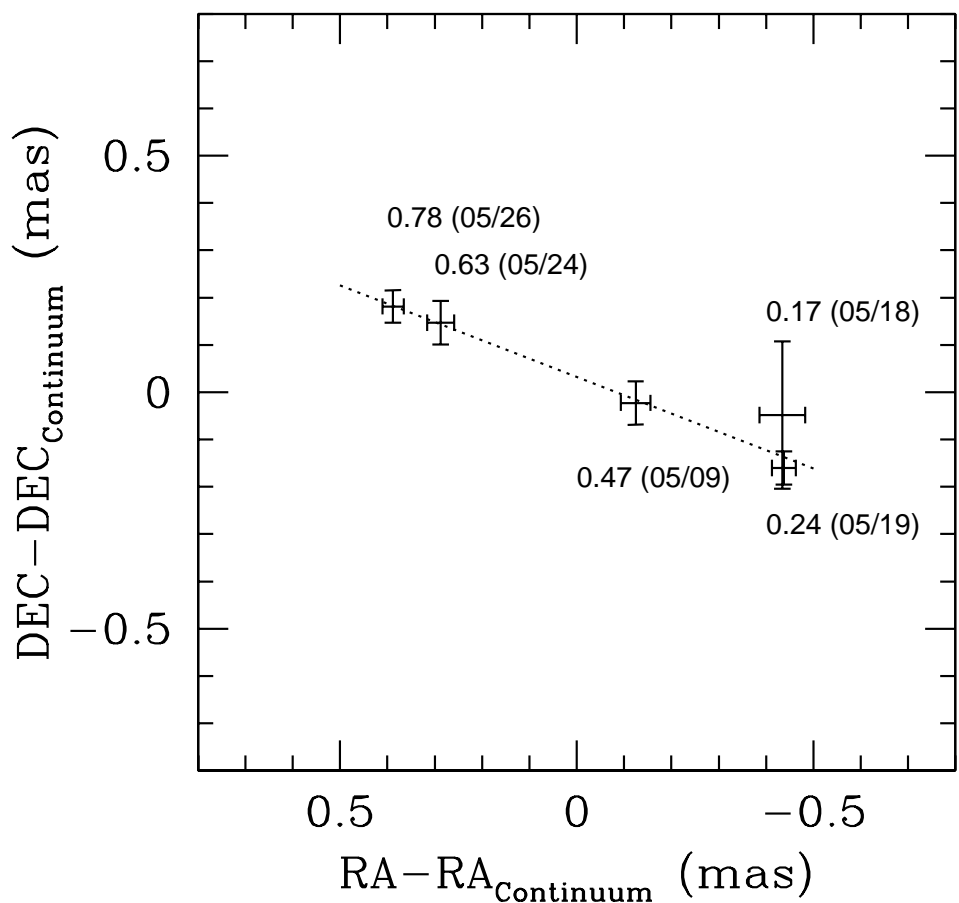}
\caption{Position of the peak of H$\alpha$ emission relative to the continuum
photocenter.
For each point we give the position uncertainty, obtained by fitting a gaussian
to the image of the source. Besides each point we also indicate the orbital phase
and the date of the observations. The dotted line was obtained by fitting a least
squares line to the data. Since $\beta$ Lyrae is an eclipsing binary, this
line is a good approximation for the orientation of the orbit in the plane
of the sky (PA=248.8$\pm1.7^{\circ}$). \label{fig9}}
\end{figure}

\begin{figure}
\plotone{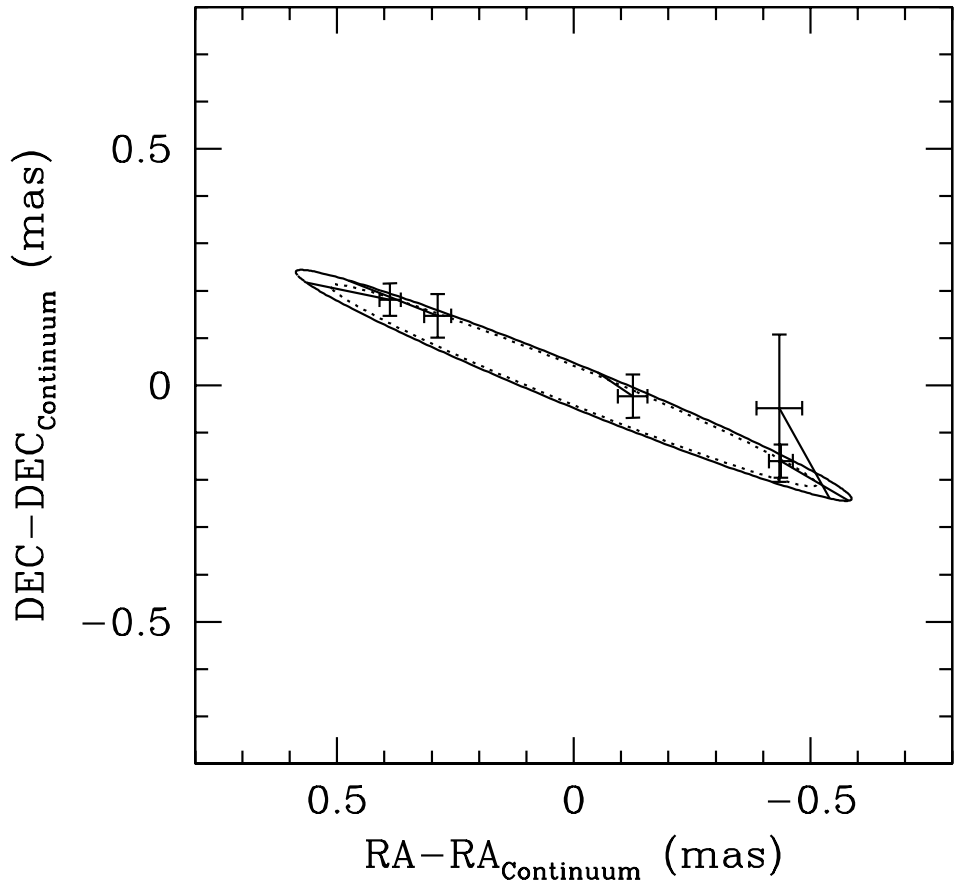}
\caption{Same as Figure 9, but displaying the orbit calculated using the
parameters described in Section~5 (solid line). The dotted line shows the
orbit obtained if we assume that the system is at a distance of 309~pc,
corresponding to a 1~$\sigma$ deviation in the distance.
The lines connecting the points to the orbit
indicate the expected position of the star in the model. \label{fig10}}
\end{figure}

\begin{figure}
\plotone{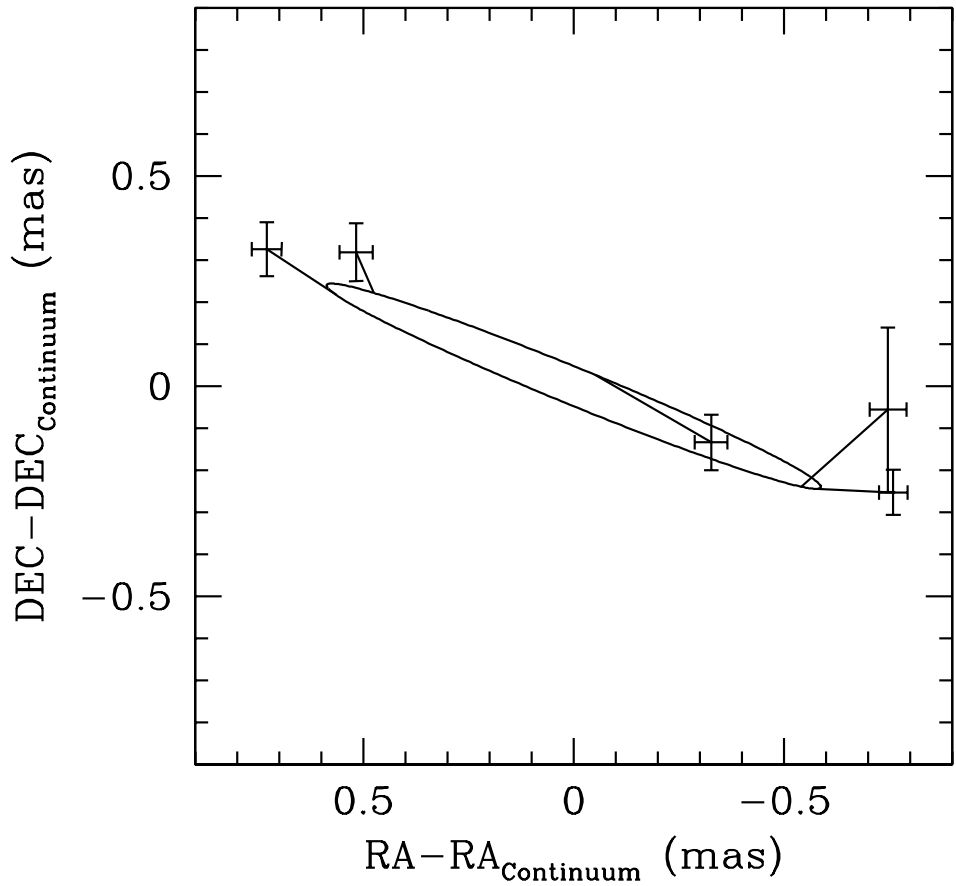}
\caption{Same as Figure 10, but displaying the H$\alpha$ positions measured
on images created using a larger correction for the continuum contribution
to the H$\alpha$ channel (c$_{cnt}=0.91$). For simplicity we show only the
model where the binary is at a distance of 270~pc. \label{fig11}}
\end{figure}

\begin{figure}
\plotone{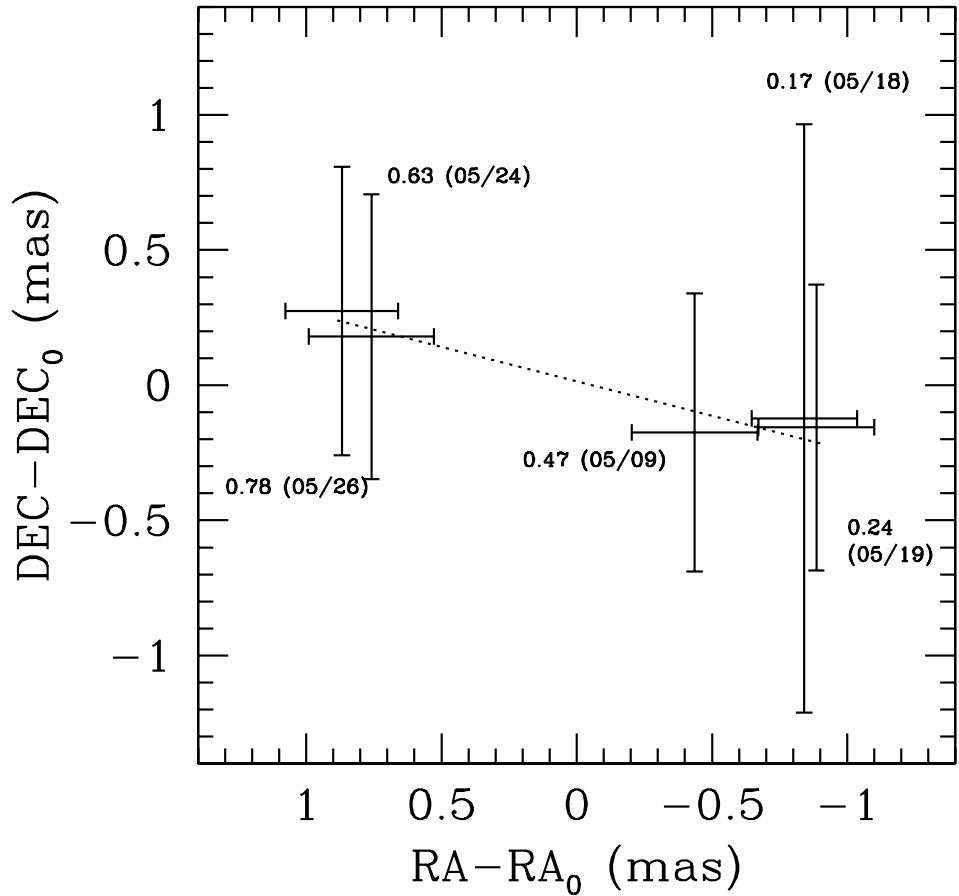}
\caption{Separation of the two binary components, obtained by fitting the
continuum V$^2$ data. For each point we give the uncertainty in the position
of the separation of the two sources. Besides each point we also indicate
the orbital phase and the date of the observations. The dotted line was
obtained by fitting a least squares line to the data, which gives
the orientation of the orbit in the plane of the sky (PA=256.1$\pm17.7^{\circ}$).
\label{fig12}}
\end{figure}

\begin{figure}
\plotone{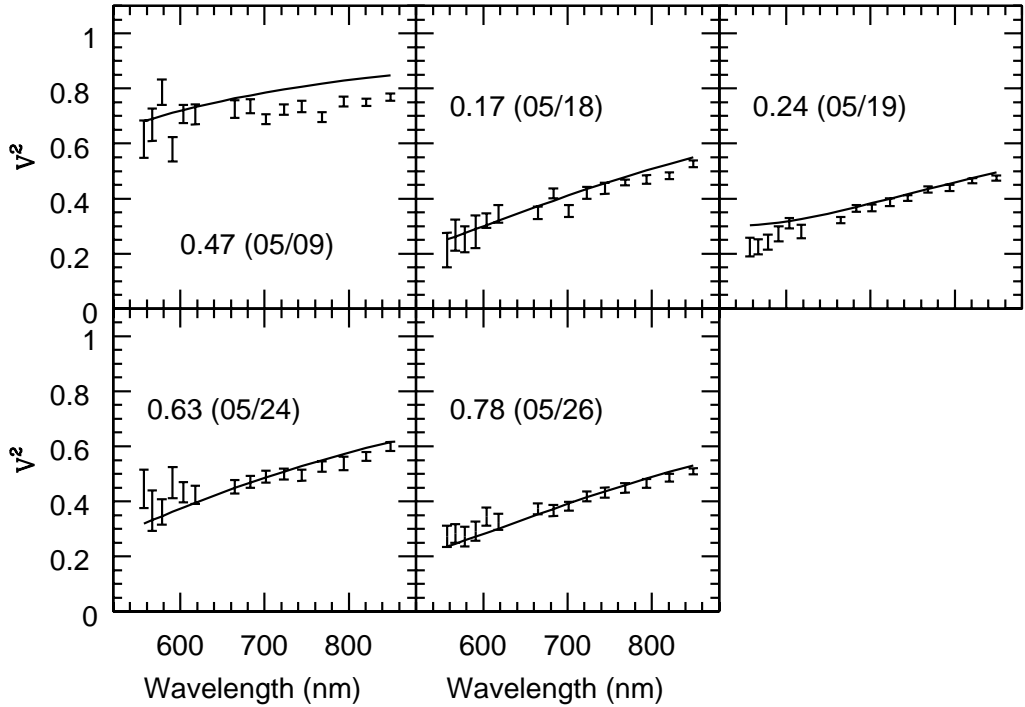}
\caption{Calibrated V$^2$'s of $\beta$ Lyrae observed with the longest baseline,
AW0-E06 (53~m). The respective nights and orbital phases are indicated
inside the panels. In order to make the visualization of the plots easier each
panel shows only one scan, corresponding to the one observed within 30 minutes of
local meridian crossing. The solid lines show the best fitting model for the
corresponding night.
\label{fig13}}
\end{figure}

\begin{figure}
\includegraphics[clip=true]{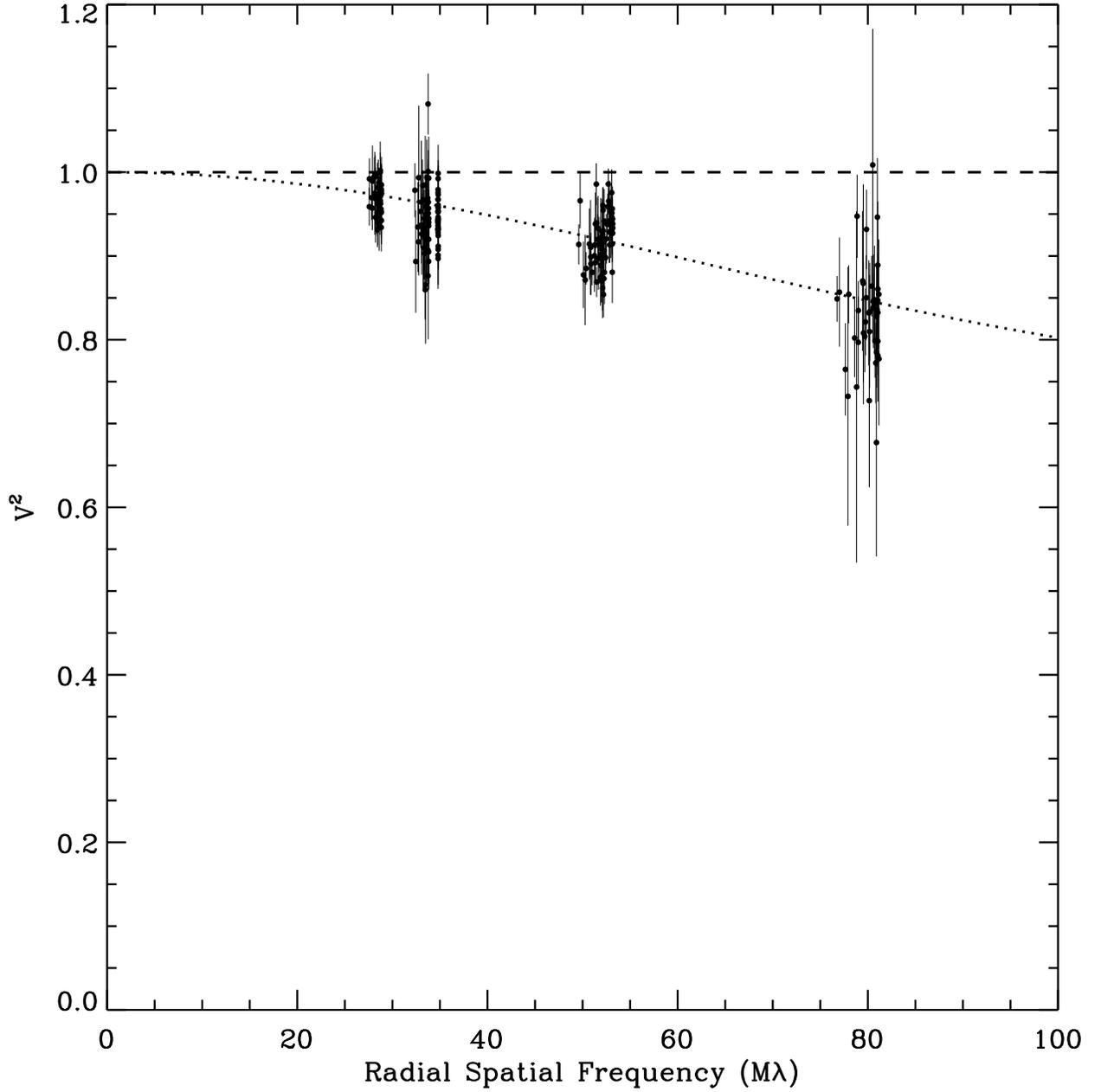}
\caption{Calibrated H$\alpha$ visibilities from the H$\alpha$ channel. The dashed
line shows the continuum V$^2$ after the removal of the binary signature. The
dotted line shows the best elliptical Gaussian model fitted to the data.
\label{fig14}}
\end{figure}

\clearpage

\begin{deluxetable}{rlrlrr}
\tabletypesize{\scriptsize}
\tablecaption{Observations Log\label{tbl-1}}
\tablewidth{0pt}
\tablehead{
\colhead{Date} & \colhead{UT Range}&\colhead{Number of Scans} & \colhead{Baselines} & \colhead{Hour Angle}&\colhead{Orbital Phase}}
\startdata
2005May09 &9.63 \ldots 11.90 &  5 & E6-AC, AW-E6, AW-AC, AE-AC, AN-AC, AE-AN & --1.50 \ldots 0.79 &0.47 \\
2005May18 &9.85 \ldots 11.80 & 10 & E6-AC, AW-E6, AW-AC, AE-AC               & --0.67 \ldots 1.28 &0.17 \\
2005May19 &9.58 \ldots 11.75 &  9 & E6-AC, AW-E6, AW-AC, AE-AC, AN-AC, AE-AN & --0.89 \ldots 1.29 &0.24 \\
2005May24 &9.08 \ldots 11.63 &  8 & E6-AC, AW-E6, AW-AC, AE-AC, AN-AC, AE-AN & --1.10 \ldots 1.48 &0.63 \\
2005May26 &8.98 \ldots 11.36 & 10 & E6-AC, AW-E6, AW-AC, AE-AC, AN-AC, AE-AN & --1.04 \ldots 1.34 &0.78 \\
\enddata
\tablecomments{The maximum lengths of the baselines shown in Column 3 are:
E6-AC 34.3~m, AW-E6 53.3~m, AW-AC 22.2~m, AE-AC 18.9~m, AN-AC 22.9~m, AE-AN 34.9~m.
We always used AC as the referenec station.}
\end{deluxetable}

\begin{deluxetable}{rrrr}
\tabletypesize{\scriptsize}
\tablecaption{Displacement of the H$\alpha$ and Continuum Photocenters\label{tbl-2}}
\tablewidth{0pt}
\tablehead{
\colhead{UT Date} & \colhead{Orbital Phase} & \colhead{RA(H$\alpha$-Continuum)} & \colhead{DEC(H$\alpha$-Continuum)}\\
\colhead{}&\colhead{}&\colhead{(mas)}&\colhead{(mas)}}
\startdata
2005May09 & 0.47 & --0.13 $\pm$ 0.03 & --0.02 $\pm$ 0.05 \\
2005May18 & 0.17 & --0.44 $\pm$ 0.05 & --0.05 $\pm$ 0.16 \\
2005May19 & 0.24 & --0.44 $\pm$ 0.03 & --0.16 $\pm$ 0.04 \\
2005May24 & 0.63 &   0.29 $\pm$ 0.03 &   0.15 $\pm$ 0.05 \\
2005May26 & 0.78 &   0.39 $\pm$ 0.02 &   0.18 $\pm$ 0.03 \\
\enddata
\tablecomments{The orbital phases were calculated based on primary eclipse ephemeris from \cite{Kreiner04}}
\end{deluxetable}

\begin{deluxetable}{rrrrrrr}
\tabletypesize{\scriptsize}
\tablecaption{Binary Separations\label{tbl-3}}
\tablewidth{0pt}
\tablehead{
\colhead{UT Date} & \colhead{Orbital Phase} & \colhead{$\rho$} & \colhead{$\theta$}
& \colhead{$\Delta_{Maj}$} & \colhead{$\Delta_{Min}$} & \colhead{$\phi$} \\
\colhead{}&\colhead{}&\colhead{(mas)}&\colhead{(deg)}&\colhead{(mas)}&\colhead{(mas)}&\colhead{(deg)}}
\startdata
2005May09 & 0.47 & 0.47 & 248.1 & 0.52 & 0.23 &   5.0 \\ 
2005May18 & 0.17 & 0.85 & 261.5 & 1.09 & 0.19 & 177.9 \\
2005May19 & 0.24 & 0.90 & 260.0 & 0.52 & 0.21 &   2.7 \\ 
2005May24 & 0.63 & 0.78 &  76.7 & 0.51 & 0.23 &   3.8  \\
2005May26 & 0.78 & 0.91 &  72.5 & 0.52 & 0.21 &   1.9 \\
\enddata
\tablecomments{Column 1: date of the observations; Column 2: orbital phase;
Columns 3 and 4: binary separation and position angle; Columns 5, 6 and 7:
position uncertainty ellipse major axis, minor axis and position angle. }
\end{deluxetable}



\begin{thebibliography}{}
\bibitem[Akeson, Swain \& Colavita (2000)]{akeson00}Akeson, R. L., Swain, M. R., \&
Colavita, M. M. 2000, \procspie, 4006, 321
\bibitem[Armstrong et al. (1998)]{Armstrong98} Armstrong, J. T., et al. 1998,
ApJ, 496, 550
\bibitem[Baldwin, et al. (1996)]{Baldwin96}Baldwin, J. E., et al. 1996, A\&A, 306, L13
\bibitem[Cot\'e \& Waters (1987)]{CW87} Cot\'e, J., \& Waters, L. B. F. M.
1987, \aap, 176, 93
\bibitem[Dobias \& Plavec (1985)]{DP85} Dobias, J.~J., \& Plavec, M.~J.
 1985, \aj, 90, 773
\bibitem[Goodricke \& Englefield (1785)]{Goodricke1785}Goodriecke, J., \&
Englefield, H. C. 1785, Philosophical Transactions Series I, 75, 153
\bibitem[Hall et al. (1994)]{Hall94} Hall, J. C., Fulton, E. E., Huenemoerder, D. P.,
Welty, A. D., \& Neff, J. E. 1994, \pasp, 106, 315
\bibitem[Harmanec (2002)]{Harmanec02}Harmanec, P. 2002, Astronomische
Nachrichten, 323, 87
\bibitem[Harmanec et al. (1996)]{Harmanec96} Harmanec, P. 1996, A\&A, 312, 879
\bibitem[Harmanec \& Scholz (1993)]{Harmanec93}Harmanec, P., \& Scholz, G.
1993, A\&A, 279, 131
\bibitem[Hoffman, Nordsieck \& Fox (1998)]{Hoffman98}Hoffman, J. L., Nordsieck,
K. H., \& Fox, G. K. 1998, AJ, 115, 1576
\bibitem[Hummel et al. (1998)]{Hummel98} Hummel, C. A., Mozurkewich, D., Armstrong,
J. T., Hajian, a. R., Elias II, N. M., \& Hutter, D. J. 1998, \aj, 116, 2536
\bibitem[Hummel et al. (2003a)]{Hummel03a} Hummel, C. A., et al. 2003a, \aj,
125, 2630
\bibitem[Hummel et al. (2003b)]{Hummel03b} Hummel, C. A., Mozurkewich, D.,
Benson, J. A., \& Wittkowski, M. 2003b, SPIE, 4838, 1107
\bibitem[Jorgensen et al. (2006)]{Jorg06} Jorgensen, A. M., et al. 2006, SPIE, 6268, 47
\bibitem[Jorgensen et al. (2007)]{Jorg07} Jorgensen, A. M., et al. 2007, \aj, 134, 1544
\bibitem[Kreiner (2004)]{Kreiner04} Kreiner, J. M. 2004, Acta Astronomica, 54, 207
\bibitem[Linnell (2000)]{Linnell00} Linnell, A. P. 2000, MNRAS, 319, 255
\bibitem[Linnell, Hubeny \& Harmanec (1998)]{Linnell98} Linnell, A. P., Hubeny, I.,
\& Harmanec, P. 1998, \apj, 509, 379
\bibitem[Monnier (2003)]{Mon03} Monnier, J. D., 2003, Reports of Progress in Physics,
66, 789
\bibitem[Monnier et al. (2007)]{Mon07} Monnier, J. D., et al. 2007, Science, 317, 342
\bibitem[Morgan, Potter \& Kondo (1974)]{Morgan74} Morgan, T. H., Potter, A. E., \&
Kondo, Y. 1974, \apj, 190, 349
\bibitem[Mourard et al. (1992)]{Mourard92} Mourard, D., et al. 1992, IAU
colloq.~135: Complementary Approaches to Double and Multiple Star Research, 32, 510
\bibitem[Owens (1967)]{owens67} Owens, J. C. 1967 \ao, 6, 51
\bibitem[Perryman et al. (1997)]{Perryman97} Perryman, M. A. C., et al. 1997,
A\&A, 323, L49
\bibitem[Peterson et al. (2006a)]{Peterson06a} Peterson, D. M., et al. 2006, ApJ,
636, 1087
\bibitem[Peterson et al. (2006b)]{Peterson06b} Peterson, D. M., et al. 2006,
Nature, 440, 896
\bibitem[Quirrenbach (1999)]{Quirrenbach99} Quirrenbach, A. 2000, in
Principles of Long Baseline Stellar Interferometry, Ed. P.~R. Lawson, p. 143 
\bibitem[Quirrenbach et al. (1994)]{Quirrenbach94} Quirrenbach, A. Buscher, D. F.,
Mozurkewich, D., Hummel, C. A., \& Armstrong, J. T. 1994, A\&A, 283, L13
\bibitem[Sahade et al. (1959)]{Sahade59}Sahade, J., Huang, S. S., Struve, O., \&
Zebergs, V. 1959, Transactions of the American Philosophical Society, 49, 32
\bibitem[Tycner et al. (2005)]{Tycner05}Tycner, C. et al. 2005, \apj, 624, 359
\bibitem[Tycner et al. (2006)]{Tycner06}Tycner, C. et al. 2006, \aj, 131, 2710
\bibitem[Umana et al. (2000)]{Umana00}Umana, G., Maxtel, P. F. L. Triglio, C.,
Fender, R. P., Leone, F., \& Yerli, S. K. 2000, A\&A, 358, 229
\bibitem[Vakili et al. (1997)]{Vakili97}Vakili, F., Mourard, D., Bonneau, D.,
Mourand, F., Stee, P. 1997, A\&A, 323, 183
\bibitem[Vakili et al. (1998)]{Vakili98} Vakili, F., et al. 1998, A\&A, 335, 261
\bibitem[van Moorsel, Kemball \& Greisen (1996)]{vanmoorsel96}van Moorsel, G.,
Kenball, A., \& Greisen, E. 1996, in ASP Conf. Ser. 101, Astronomical Data
Analysis Software and Systems V, ed. G. H. Jacoby \& J. Barnes (San Francisco:ASP), 37
\bibitem[Wilson (1974)]{Wilson74}Wilson, R. E. 1974, \apj, 189, 319
\bibitem[Yoon et al. (2007)]{Yoon07} Yoon, J. et al. 2007, PASP, 119, 437
\end{thebibliography}
\end{document}